\documentclass{iopart}
\usepackage{graphicx}  
\usepackage{latexsym}  

\jl{6}        
\eqnobysec    

\def\beq{\begin{equation}}
\def\eeq{\end{equation}}
\def\rmd{{\rm d}}
\def\version{\today}

\def\zero{{}^{(0)}\kern-1pt}
\def\one{{}^{(1)}\kern-1pt}
\def\two{{}^{(2)}\kern-1pt}

\begin{document}

\begin{flushright}
Current version: \version \\
\end{flushright}

\title[Kerr metric, static observers and Fermi coordinates]
{Kerr metric, static observers and Fermi coordinates}

\author{
Donato Bini$^*{}^\S$,
Andrea Geralico${}^\S$ and
Robert T Jantzen${}^\S\,{}^\diamond$
}
\address{
  ${}^*$\
Istituto per le Applicazioni del Calcolo ``M. Picone'', CNR I-00161 Rome, Italy
}
\address{
  ${}^\S$\
  International Center for Relativistic Astrophysics,
  University of Rome, I-00185 Rome, Italy
}
\address{${}^\diamond$
Department of Mathematical Sciences, Villanova University, Villanova, PA 19085, USA
}

\begin{abstract}
The coordinate transformation which maps the Kerr metric written in standard Boyer-Lindquist coordinates to its corresponding form 
adapted to the natural local coordinates of an observer at rest at a fixed position in the equatorial plane, i.e., Fermi coordinates for the  neighborhood of a static observer world line, is derived and discussed in a way which extends to any uniformly circularly orbiting observer there. 
%
\end{abstract}

\pacno{04.20.Cv}

\section{Introduction}

For a timelike world line of an observer in flat or curved spacetime, Fermi coordinates \cite{fermi,bobfermi,manasse}
(so named by Synge \cite{synge})
are one of the most natural generalizations of the inertial Cartesian coordinates of a single inertial observer in flat spacetime, and the axes associated with these coordinates form the observer's \lq\lq proper reference frame" discussed at length by Misner, Thorne and Wheeler \cite{MTW}. 
It can be shown that there always exist Fermi coordinates $(X^A)=(T,X^i)$ ($A=0,1,2,3$, $i=1,2,3$), adapted to the world line of an accelerated observer and covering a small spacetime region around it, in terms of which the spacetime metric can be expressed in the form 
\begin{eqnarray}
\label{fermimetric}
\fl\quad
\rmd s^2=-(1+2A_i(T)X^i)\rmd T^2 - 2\epsilon_{ijk}X^i \omega^{j}(T) \rmd X^k dT 
          + \delta_{ij} \rmd X^i \rmd X^j + O(3)
\end{eqnarray}
through terms of first order in the spatial Fermi coordinates (see Eq.~(13.71) of \cite{MTW}). 
The observer is at rest at the spatial origin $X^i=0$ of these coordinates and $T$ represents the proper time along the observer world line. $A_i(T)$ are the coordinate components of the observer acceleration along the world line, 
while $\omega^i(T)$ is the angular velocity of the axes aligned with the Fermi spatial coordinates with respect to a Fermi-Walker transported frame along the world line. The $X^i$ coordinate lines are spacelike geodesics orthogonal to the observer world line, and the full Fermi coordinate frame is characterized by the vanishing of all the connection coefficients $\Gamma^A{}_{BC}$ along the given world line except for those mixed space and time components corresponding to the acceleration and angular velocity, and the first order Fermi metric components are expressible entirely in terms of these coefficients, leading to the above simple form.

For a world line which is a Killing trajectory in a spacetime with symmetry, 
one can choose the Fermi spatial frame to be Lie dragged along the world line, so that
these quantities are time-independent. The Fermi-Walker angular velocity is then just the vorticity of the corresponding Killing congruence \cite{vish}, which follows directly from the interpretation of the vorticity as the local rotation of nearby world lines in the congruence (since Lie dragging locks the Fermi spatial frame to connecting vectors).

One can align the acceleration with the first spatial axis $A=A^1$. 
For the circular Killing orbits considered here, the angular velocity is orthogonal to the acceleration and can be aligned with the second spatial axis $\Omega=\omega^{\hat 2}$, leading to the specialized form
\begin{eqnarray}
\label{fermimetric2}
\fl\quad
\rmd s^2=-(1+2A X^1)\rmd T^2+ 2\Omega (X^3 \rmd X^1- X^1 \rmd X^3 )\rmd T  
            + \delta_{ij} \rmd X^i \rmd X^j + O(3)\ .
\end{eqnarray}

One is still confronted with the problem of finding the transformation of coordinates from those which describe the given spacetime of interest to the Fermi coordinates, if one wants to use relate Fermi coordinate calculations to the original coordinates. This explains locally the correct interpretation of the deflection of light in the uniform gravitational field at the surface of the Earth, for example, neglecting its rotation, as discussed by Linet \cite{linet}, where naively introduced local Cartesian coordinates can be misleading \cite{diaz}.

Here we first discuss the explicit coordinate transformation mapping the Kerr metric written in standard Boyer-Lindquist coordinates 
$(x^\alpha)=(t,r,\theta,\phi)$ ($\alpha=0,1,2,3$),
to its corresponding form (\ref{fermimetric2}) along the world line of a static observer at rest at a fixed position on the equatorial plane, up to second order in the spatial Fermi coordinates $X^i$, for a neighborhood of a static observer located at any point on the equatorial plane, generalizing previous results by Linet \cite{linet,linet2} valid in the case of the Schwarzschild solution. This gives the Fermi metric up to first order in the spatial Fermi coordinates.
With a little more effort, using comoving coordinates, the same technique can be used to obtain the corresponding Fermi coordinates along any uniformly rotating circular orbit in the equatorial plane. From the geometrical interpretation of the second order coefficients in the Kerr case, ones sees the more general result typical of any timelike circular orbit in an axially symmetric stationary spacetime in relation to an adapted system of comoving coordinates, provided that only first derivatives of the metric with respect to one spatial coordinate are nonzero. For Kerr
one has the extra discrete symmetry characterizing the equatorial plane ($\theta =\pi/2$), namely that it is invariant under the transformation $\theta \to \pi-\theta$ so that all first order $\theta$ derivatives of the metric vanish there $g_{\alpha\beta,\theta}\vert_{\theta=\pi/2}=0$, simplifying the problem so that only $r$ derivatives need be considered for the second order expansion of the coordinate transformation. Similarly one can apply these results to circular orbits in the G\"odel or Minkowski spacetime in cylindrical coordinates. For this class of spacetimes the Frenet-Serret angular velocity 
has only a single nonzero component along the $\theta$ direction. Modulo the sign which depends on convention, this is just the Frenet-Serret first torsion $\tau_1$ of the world line,  \cite{circfs,fsfw,idcf1,idcf2}, important in the calculation of the precession of Fermi-Walker transported spin vector of a test gyroscope.

Of course the fact that the spatial Boyer-Lindquist coordinates are orthogonal and hence the spatial metric in the local rest spaces of the static observers is diagonal considerably simplifies the problem, leaving only the tilting of the local rest spaces in the azimuthal direction to contribute nondiagonal terms. Thus only the intrinsic spatial curvatures along the spatial coordinate axes together with the velocity describing the boost of the local rest space of the static observers (and its gradient) enter at the first derivative level, and only radial first derivatives enter in the equatorial plane because of the reflection symmetry across it. This combination of circumstances makes the problem feasible to solve and the results rather straightforward to interpret. To first order the coordinate transformation must contain the various terms which, when combined with the first derivatives of the Boyer-Lindquist metric, lead to the first order terms in the Fermi metric which are expressible entirely in terms of the Fermi coordinate connection coefficients, all of which vanish except for the kinematical terms.

None of these results would have been possible without the combination of insightful choice and identification of variables with the power of a computer algebra system, in our case Maple. In fact having understood the process, one can continue the coordinate transformation to the third order in the Fermi coordinates where the curvature enters, necessary to transform between the Boyer-Lindquist line element and the well known form of the Fermi metric expressed to second order in the Fermi coodinates, found for geodesics by Manasse and Misner \cite{manasse} and extended to arbitrary acceleration and Fermi-Walker angular velocity by Ni and Zimmerman \cite{ni}. However, now the third order coefficients in the coordinate transformation must involve many different kinds of terms which contribute to spacetime curvature when properly combined with other second derivatives of the metric in the expansion process of the metric transformation, since the second order coefficients in the Fermi metric are all directly expressible in terms of the Riemann tensor components along the observer world line, and hence since the spacetime is a vacuum spacetime, in terms of the electric and magnetic parts of the Weyl tensor. The third order coefficients are thus much more complicated in form and origin and do not lend themselves so easily to interpretation as in the second order case.

Fukushima has derived the second order approximation to Fermi coordinates for a specified world line in the nonrotating post-Newtonian metric in post-Newtonian quasi-Cartesian coordinates \cite{fuk}. Such considerations are important in the more complicated arena of celestial coordinate reference systems, astrometry and geodesy \cite{kop}.

Of course in the Kerr spacetime one can do local calculations entirely in Fermi coordinates without knowing the transformation of coordinates from those in which the original world line and quantities along it are expressed, as is the case in evaluating tidal deformations along timelike world lines in black hole spacetimes.
Marck \cite{marck} showed how to work with the Fermi metric alone along geodesics in Kerr to study this question, recently revisited by Ishii, Shibata and Mino \cite{ism}.
Chicone and Mashhoon \cite{chicmash02,chicmash04,chicmash05_1,chicmash05} have extended the discussion to more general accelerated world lines (see also \cite{bertotti,nesterov,carneiro,rindler} and references therein).
However, because of the importance of the local observer frame in gravitational calculations, it is satisfying to finally understand how Fermi coordinates may be explicitly constructed for interesting world lines in the complicated geometry of the Kerr spacetime. 

\section{Fermi coordinates for static observers in Kerr}

The Kerr metric in standard Boyer-Lindquist coordinates $(x^\alpha) =(t,r,\theta,\phi)$ is given by
\begin{eqnarray}
\label{kerrmetric}
\fl\quad
\rmd s^2 
=  g_{\alpha\beta} \rmd x^\alpha  \rmd x^\beta 
&=& -\left(1-\frac{2\mathcal{M}r}{\Sigma}\right)\rmd t^2 
-\frac{4a\mathcal{M}r}{\Sigma}\sin^2\theta\rmd t\rmd\phi
+\frac{\Sigma}{\Delta}\rmd r^2 
+\Sigma\rmd \theta^2\nonumber\\
\fl\quad
&&+\frac{(r^2+a^2)^2-\Delta a^2\sin^2\theta}{\Sigma}\sin^2 \theta \rmd \phi^2\ ,
\end{eqnarray}
where $\Delta=r^2-2\mathcal{M}r+a^2$ and $\Sigma=r^2+a^2\cos^2\theta$. Here $\mathcal{M}$ and $a$ are the total mass and specific angular momentum characterizing the spacetime. 

To obtain the metric (\ref{fermimetric2}) up to first order in the Fermi coordinates, one needs the coordinate transformation up to second order. This involves 16 first order coefficients $a^\alpha{}_B$ and $4\times 10$ second order coefficients $b^\alpha{}_{BC}$ 
\beq
x^\alpha = x^\alpha_0 + a^\alpha{}_B X^B + b^\alpha{}_{BC} X^B X^C + O(3)\ ,
\eeq
with Jacobian
\beq
    J^\alpha{}_B = a^\alpha{}_B  + 2 b^\alpha{}_{BC} X^C + O(2)\ ,
\eeq
which must satisfy the 10 equations
\beq\label{eq:gJg}
  g_{AB} =    g_{\alpha\beta} J^\alpha{}_A J^\beta{}_B
 = a^\alpha{}_A a^\beta{}_B + 2 g_{\alpha\beta} a^\alpha{}_A b^\beta{}_{BC}X^C
           + O(2)\ .
\eeq
The first order coefficients are completely determined by aligning the Fermi coordinate axes with a preferred orthonormal frame adapted to these time coordinate world lines, a natural choice for which is a Lie dragged frame along the given Killing trajectory.
Then once the Boyer-Lindquist and Fermi metric components contained in Eqs.~(\ref{eq:gJg}) are both expanded up to the appropriate order in the Fermi coordinates, dropping second order terms, these become 10 linear equations in the four Fermi coordinate variables which must be identically satisfied, giving exactly 40 linear equations for the 40 second order coefficients. 

For the special case of a Killing trajectory world line like the time coordinate world lines representing points \lq\lq fixed in space," one can basically work at a single point along the world line, since the Fermi coordinates associated with an invariant frame will respect this symmetry. 
Note that since the equatorial plane $\theta=\pi/2$ is invariant under the reflection symmetry of the metric $\theta \to \pi-\theta$, the derivatives $g_{\alpha\beta,\theta}$ vanish there, and hence expansions of the background metric involving these derivatives must vanish too. Thus only the radial derivatives contribute when expanding the background metric, since the metric only depends on $r$ and $\theta$ due to its stationary axially symmetric form.
Along the world line, the Fermi coordinate frame reduces to the chosen Lie dragged comoving frame from which it was built.

In the equatorial plane of the Kerr spacetime where the time lines are radially accelerated (true for any uniformly circularly rotating observer), one can use the unit vectors along the $t$, $r$, $\theta$, and boosted $\phi$ coordinate directions (orthogonal to the time coordinate lines) as the Fermi frame along the time coordinate world line, with the  $X^1$-axis aligned with the radial coordinate $r$ and $T$, $X^2$ and $X^3$ respectively with the $t$, $\theta$ and boosted $\phi$ directions, leaving only the second order coefficients to be determined, subject to the condition that only $t$ change when the spatial Fermi coordinates are zero and $T$ varies (so that the origin of coordinates represents that given world line), eliminating any quadratic terms in $T$ from the representation of the spatial Boyer-Lindquist coordinates. 

To start this procedure, one must introduce the stationary axially symmetric orthonormal threading frame adapted to the static observers
\beq\fl
e_{\hat t} =M^{-1} \partial_t \equiv m\ , \,\
e_{\hat r} =\gamma_{rr}^{-1/2} \partial_r\ , \,\
e_{\hat \theta} =\gamma_{\theta\theta}^{-1/2} \partial_\theta\ , \,\
e_{\hat \phi} = \gamma_{\phi\phi}^{-1/2} (\partial_\phi  + M_\phi \partial_t )\ ,
\eeq
with dual frame
\beq\fl\qquad
\omega^{{\hat t}} = M (\rmd t - M_\phi \rmd\phi)\ , \ 
\omega^{{\hat r}} =  \gamma_{rr}^{1/2} \rmd r\ , \ 
\omega^{{\hat \theta}} = \gamma_{\theta\theta}^{1/2} \rmd \theta\ , \
\omega^{{\hat \phi}} = \gamma_{\phi\phi}^{1/2} \rmd\phi \ ,
\eeq
where $m$ denotes the 4-velocity of the static observers.
This leads to the so-called threading form for the spacetime metric
\beq
\rmd s^2 = -M^2 (\rmd t - M_\phi \rmd\phi)^2 + \gamma_{rr} \rmd r^2 + \gamma_{\theta\theta}\rmd \theta^2 + \gamma_{\phi\phi} \rmd\phi^2 \ .
\eeq
Explicitly
\begin{eqnarray}
[M,M_\phi] &=& [(1-2\mathcal{M}/r)^{1/2}, -2a\mathcal{M}/(r-2\mathcal{M})]\ ,\nonumber\\\relax
[\gamma_{rr}^{1/2},\gamma_{\theta\theta}^{1/2},\gamma_{\phi\phi}^{1/2}]
&=& [r/\Delta^{1/2}, \Delta^{1/2} , (r\Delta/(r-2\mathcal{M}))^{1/2}]\ .
\end{eqnarray}

The orthonormalization of the coordinate frame in two steps (diagonalization and normalization) 
\beq
\omega^{{\hat\alpha}}= \omega^{{\hat\alpha}}{}_\beta \rmd x^{\beta}\ ,\qquad
e_{{\hat\alpha}}= e^\beta{}_{{\hat\alpha}}\partial_\beta
\eeq
defines two matrices
\beq
(\omega^{{\hat\alpha}}{}_\beta)
= \left(\begin{array}{cccc} 1&0&0&\nu_{(n)}\\ 0&1&0&0\\ 0&0&1&0\\ 0&0&0&1\end{array}\right)
  \left(\begin{array}{cccc} M&0&0&0\\ 0&\gamma_{rr}^{1/2}&0&0\\ 0&0&\gamma_{\theta\theta}^{1/2}&0\\                        
                    0&0&0&\gamma_{\phi\phi}^{1/2}\end{array}\right)\ ,
\eeq
where
\beq
\nu_{(n)} = \nu(n,m) = -\gamma_{\phi\phi}^{-1/2}M M_\phi
  = 2aM/(r\Delta^{1/2})
\eeq
is the relative velocity with respect to the static observers of the observers moving normal to the time hypersurfaces (ZAMOs = zero angular momentum observers) with 4-velocity $n$. These matrices depend only on $r$ in the equatorial plane, and only their $r$ derivatives are needed for our calculations. The derivative of the relative velocity can be expressed in terms of the vorticity, whose only nonvanishing component is orthogonal to the equatorial plane
\beq
2\omega(m)^{\hat\theta}
= -\gamma_{\hat\phi\hat\phi}^{-1/2} M M_{\phi,{\hat r}}
= \nu_{(n)}{}_{,\hat r} - \nu_{(n)} a(m)^{\hat r} - \nu_{(n)} \kappa(\phi,m)^{\hat r}\ .
\eeq

Similarly one can introduce the (threading, i.e., with the timelike axis along the $T$ coordinate lines) orthonormal frame associated with the Fermi coordinates 
$ W^{{\hat\alpha}} = W^{{\hat\alpha}}{}_B \rmd X^B$,
with (to first order)
\beq\fl\qquad
(W^{{\hat\alpha}}{}_B) 
= \left(\begin{array}{cccc} 1&-\Omega X^3&0&\Omega X^1\\ 0&1&0&0\\ 0&0&1&0\\ 0&0&0&1\end{array}\right)
  \left(\begin{array}{cccc} 1+AX^1&0&0&0\\ 0&1&0&0\\ 0&0&1&0\\  0&0&0&1\\ 
   \end{array}\right)\ .
\eeq
Although the Fermi coordinates are orthogonal along the observer world line, away from the spatial origin, the Fermi coordinate time lines are no longer orthogonal to the three orthogonal spatial coordinate directions, so in orthogonalizing the Fermi coordinate frame, one must either start with the time direction (threading point of view) or the three spatial coordinate directions (slicing point of view).
It is also useful to remember the index correspondence $(X^1,X^2,X^3)\sim (r,\theta,\phi)$ along the chosen world line.

For the world line of a static observer located at $r=r_0$, $\theta=\theta_0=\pi/2$, $\phi=\phi_0$, 
the relationship between the starting coordinates and the Fermi coordinates can be taken to have the form
\beq\label{eq:omegadeltax}
 \omega^{{\hat\alpha}}{}_\beta \vert_{X^i=0} (x^\alpha - x^\alpha_0) 
 = \delta^{{\hat\alpha}}{}_B X^B +  b^{{\hat\alpha}}_{BC} X^B X^C  + O(3)\ ,
\eeq
so that by exterior differentiation
\beq
 \zero\omega^{{\hat\alpha}} \equiv  \omega^{{\hat\alpha}} \vert_{X^i=0} 
 =  [\delta^{{\hat\alpha}}{}_B + 2 b^{\hat\alpha}_{0B} X^0  ] \rmd X^B  + O(2)\ ,
\eeq
and hence the coordinate transformation Jacobian is
\beq
J^\alpha{}_B = \partial x^\alpha/\partial X^B 
    = \zero e^\alpha{}_{\hat\gamma}  [\delta^{\hat\gamma}{}_B + 2 b^{\hat\gamma}{}_{BC}X^C   ]   + O(2)\ ,
\eeq
The left superscript notation $\zero f= f\vert_{X^i=0}$ for evaluation of quantities along the given observer world line (at fixed $r_0$, $\theta_0$) is useful to avoid index confusion on the right.

For $ \zero \omega^{{\hat\alpha}}$ to agree with $ \zero W^{{\hat\alpha}} = \delta^{{\hat\alpha}}{}_B \rmd X^B $ along the world line as assumed,
one must have $b^{{\hat\alpha}}_{0B}=0$, which means that the right hand side of 
Eq. (\ref{eq:omegadeltax}) is independent of the Fermi time $T$,
so at most the $24=4\times6$ coefficients $b^{\hat\alpha}_{ij}$ will be nonzero.
This also guarantees that the spatial origin of coordinates $X^i=0$ coincides with the given world line, equivalent to $b^{{\hat i}}_{00}=0$. 

The coordinate transformation of the metric is then
\beq\label{gewwJJ}
  g_{AB} = \eta_{\hat\gamma\hat\delta} \omega^{\hat\gamma}{}_\alpha  \omega^{\hat\delta}{}_\beta
             J^\alpha{}_A J^\beta{}_B\ .
\eeq
To expand the right hand side to first order, one needs to expand the background frame to first order
\begin{eqnarray}
\omega^{\hat\gamma}{}_\alpha  
&=& \zero \omega^{\hat\gamma}{}_\alpha 
    + \zero(\omega^{\hat\gamma}{}_{\alpha,{\hat r}})  X^1  + O(2)
    \nonumber\\
&=& \left[\delta^{\hat\gamma}{}_\delta + \zero (\omega^{\hat\delta}{}_{\mu,{\hat r}}  
            e^\mu{}_{\hat\delta})  X^1 \right]
    \zero \omega^{\hat\delta}{}_{\alpha}   + O(2)  \ .
\end{eqnarray}

The $\hat r$ derivatives of the static observer orthonormal frame components are therefore best evaluated in the matrix form
\beq
  (\omega^{\hat\alpha}{}_{\gamma,{\hat r}}  e^\gamma{}_{\hat\beta})
 = {\rm diag}(a(m)_{\hat r}, -\kappa(x^i,m)_{\hat r})
    + (2\omega(m)^{\hat\theta} \delta^{\hat\alpha}{}_0  \delta^{\hat\phi}{}_{\hat\beta}) 
      \ ,
\eeq
where $a(m)^{\hat r}=(\ln M)_{,\hat r}$ is the radial acceleration of the static observers, 
$\omega(m)^{\hat\theta}$
is the only nonzero component of the vorticity of the static observers, 
and $\kappa(x^i,m)^{\hat r}=-(\ln  \gamma_{ii}^{1/2})_{,\hat r}$ are the radial components of the curvature vectors associated with the diagonal metric coefficients
\beq
 a(m)^{\hat r} = \frac{\mathcal{M}\Delta^{1/2}}{r^2(r-2\mathcal{M})}\ ,\
 \omega(m)^{\hat\theta} = -\frac{a\mathcal{M}}{r^2(r-2\mathcal{M})}\ ,\
\eeq
\beq\fl
 \kappa(r,m)^{\hat r} = \frac{Mr-a^2}{r^2\Delta^{1/2}}\ ,\
 \kappa(\theta,m)^{\hat r} = -\frac{\Delta^{1/2} }{r^2}\ ,\
 \kappa(\phi,m)^{\hat r} = -\frac{r(r-2\mathcal{M})^2- \mathcal{M}a^2}{r^2\Delta^{1/2}(r-2\mathcal{M})} \ .
\eeq
These five spatial quantities are just the nonvanishing orthonormal components of the frame transformation matrix derivatives which represent the inhomogeneous terms necessary for the transformation from the Christoffel symbols of the Boyer-Lindquist metric to those of the Fermi metric.

The first order part of Eq.~(\ref{gewwJJ}) is then
\begin{eqnarray}\fl
&&  2 A X^1 \delta^0{}_A \delta^0{}_B  
  + 2\Omega (X^3 \delta^1{}_{(A} X^1 \delta^3{}_{(A} )\delta^0{}_{B)} 
\nonumber\\ &&\qquad
= 2 \eta_{\hat\gamma\hat\delta} 
  \,\zero (\omega^{\hat\gamma}{}_{\mu,{\hat r}}  
            e^\mu{}_{\hat\alpha}) X^1 \delta^{\hat\alpha}{}_A   \delta^{\hat\delta}{}_B
            + 4  \eta_{\hat\gamma\hat\delta} b^{\hat\gamma}{}_{Aj}\delta^{\hat\delta}{}_B X^j
\ ,
\end{eqnarray}
Equating the coefficients of the four Fermi coordinates on both sides of these ten equations, and solving these 40 linear equations for the 40 second order coefficients as well as $A$ and $\Omega$, one recovers the expected results
$A= a(m)^{\hat r}$ and $\Omega=\omega^2 =\omega(m)^{\hat\theta}$ for the first order coefficients, while expressing the second order coefficients entirely in terms of the angular velocity and curvatures. 
With hindsight looking at the solutions below, the frame transformation matrix derivative terms that enter into the expansion coefficients in the Jacobian factors in Eq.~(\ref{gewwJJ}) and the first derivatives of the Boyer-Lindquist metric factors (again in the form of frame transformation derivative terms) on the right hand side both contribute to the coordinate transformation at this order so that when combined in the expansion process, they form combinations of the Christoffel symbols of the Fermi metric, in terms of which the first order terms on the left hand side can be expressed, and all of which vanish along the observer world line except for the acceleration and angular velocity terms.

The relationship between the Boyer-Lindquist coordinates ($t, r, \theta, \phi$) and the Fermi coordinates ($T,X^i$) is finally found to be
\begin{eqnarray}
\fl \qquad
\zero M[(t-t_0)-\zero M_\phi (\phi-\phi_0)]&=& T 
    -\zero \omega(m)^{\hat\theta} X^1 X^3
     + O(3)
\ , \nonumber\\\fl \qquad
\phantom{-\zero M_\phi (\phi-\phi_0)}\zero \gamma_{rr}^{1/2}(r-r_0)
&=& X^1(1+ {\textstyle \frac12}\zero \kappa(r,m)^{\hat r}X^1)         
 -{\textstyle \frac12} \zero \kappa(\theta,m)^{\hat r} (X^2)^2 
\nonumber \\\fl \qquad &&
          -{\textstyle \frac12} \zero \kappa(\phi,m)^{\hat r} (X^3)^2
           + O(3)
\ ,\nonumber \\\fl \qquad
\phantom{-\zero M_\phi (\phi-\phi_0)}\zero \gamma_{\theta\theta}^{1/2}(\theta-\theta_0)
&=& X^2(1+ \zero \kappa(\theta,m)^{\hat r}X^1)
 + O(3)
\ ,\nonumber \\\fl \qquad
\phantom{-\zero M_\phi (\phi-\phi_0)}\zero \gamma_{\phi\phi}^{1/2}(\phi-\phi_0)
&=& X^3(1+ \zero \kappa(\phi,m)^{\hat r}X^1)
 + O(3)
\ .
\end{eqnarray} 
The terms of the form \lq\lq $\kappa X^1$" result from the variation of the diagonal metric coefficients with radius, while the vorticity term comes from the variation of the relative velocity, and the additional curvature terms involving $X^2$ and $X^3$ in the radial coordinate transformation are just the quadratic approximation of the \lq\lq spherical" $r$ coordinate surfaces with respect to the geodesic plane $X^1=0$ in the local rest space of the static observer. 

The extra factor of $\frac12$ for the $(X^1)^2$ radial curvature correction term compared to the corresponding angular correction factors is easily understood. Along an $X^1$ coordinate line from the origin of coordinates, along which $\gamma_{rr}^{1/2} dr = \rmd X^1 + O(2)$ holds, one has
\begin{eqnarray}\fl\qquad
r-r_0  &=& \int \rmd r 
 = \int [\gamma_{rr}^{-1/2} +O(2)] \rmd X^1 
\nonumber\\\fl\qquad
 &=& \zero \gamma_{rr}^{-1/2} \int [1+ \zero \kappa(r,m)^{\hat r} X^1] \rmd X^1
+O(3)\ ,
\end{eqnarray} 
from which integration yields
\beq
\zero \gamma_{rr}^{1/2}  (r-r_0) = X^1 + {\textstyle\frac12}\zero \kappa(r,m)^{\hat r} (X^1)^2
+O(3)\ .
\eeq
Instead doing a similar calculation along an $X^2$ or $X^3$ coordinate line at constant but nonzero $X^1$, then $X^1$ is a constant in the integral so no factor of $\frac12$ arises.
These curvature correction terms, together with the metric correction terms, are responsible for the difference between naively defined local Cartesian coordinates about the given world line 
and the correct Fermi coordinates pointed out by Linet \cite{linet2}.

Of course then one must remove the frame transformation (multiplying by its inverse) to get the final transformation of coordinates, re-expressing the additional $\phi$ term on the left hand side of the first equation in terms of the Fermi coordinates and moving it to the right hand side
\begin{eqnarray}
\fl \quad
t-t_0&=& \zero M^{-1}[T 
 -\zero \nu(n,m)X^3(1+ \zero \kappa(\theta,m)^{\hat r}X^1) 
     -\zero \omega(m)^{\hat\theta} X^1 X^3]
     + O(3)
\ , \nonumber\\\fl \quad
r-r_0
&=& \zero \gamma_{rr}^{-1/2} X^1(1+ {\textstyle \frac12}\zero \kappa(r,m)^{\hat r}X^1)
\ ,\nonumber \\\fl \qquad &&
          -{\textstyle \frac12} \zero \kappa(\theta,m)^{\hat r} (X^2)^2 
          -{\textstyle \frac12} \zero \kappa(\phi,m)^{\hat r} (X^3)^2
           + O(3)
\ ,\nonumber \\\fl \quad
\theta-\theta_0
&=& \zero \gamma_{\theta\theta}^{-1/2} X^2(1+ \zero \kappa(\theta,m)^{\hat r}X^1)
 + O(3)
\ ,\nonumber \\\fl \quad
\phi-\phi_0
&=& \zero \gamma_{\phi\phi}^{-1/2} X^3(1+ \zero \kappa(\phi,m)^{\hat r}X^1)
 + O(3)
\ .
\end{eqnarray} 
The first equation shows the boost necessary between the two time coordinates with relative velocity $\zero \nu(n,m)$, as well as the additional second order desynchronization of the Fermi time coordinate  
due to the vorticity. Both these quantities vanish in the Schwarzschild limit $a=0$, for which the coordinate transformation reduces to the one given by Linet \cite{linet}, but with a geometrical interpretation of the coefficients which explains their origin.

\section{Zero Angular Momentum Observers}

The same technique may be used to find the Fermi coordinates along the zero angular momentum observers (ZAMOs), provided one introduces the natural orthonormal frame associated with their decomposition of the tangent space orthogonal to the slicing by the $t$ coordinate hypersurfaces. Letting $n$ be the ZAMO 4-velocity, in the equatorial plane this frame is 
\beq\fl
e_{\hat t} =N^{-1} (\partial_t -N^\phi\partial_\phi )\equiv n\ , \,\
e_{\hat r} =g _{rr}^{-1/2} \partial_r\ , \,\
e_{\hat \theta} =g _{\theta\theta}^{-1/2} \partial_\theta\ , \,\
e_{\hat \phi} = g _{\phi\phi}^{-1/2} \partial_\phi\ ,
\eeq
with dual frame
\beq\fl\qquad
\omega^{{\hat t}} = N \rmd t \ , \ 
\omega^{{\hat r}} =  g _{rr}^{1/2} \rmd r\ , \ 
\omega^{{\hat \theta}} = g _{\theta\theta}^{1/2} \rmd \theta\ , \
\omega^{{\hat \phi}} = g _{\phi\phi}^{1/2}( \rmd\phi +N^\phi \rmd t )\ ,
\eeq
in terms of which the metric takes the so-called slicing form
\beq
\rmd s^2 = - N^2 \rmd t^2  +  g _{rr}\rmd r^2  + g _{\theta\theta}\rmd \theta^2 + g _{\phi\phi}(\rmd\phi +N^\phi \rmd t )^2
\ .
\eeq
Explicitly
\begin{eqnarray}
[N,N_\phi] &=& [[r\Delta/(r^3+a^2r+2a^2M)]^{1/2}, -2aM/r]\ ,\nonumber\\\relax
[g _{rr}^{1/2},g _{\theta\theta}^{1/2},g _{\phi\phi}^{1/2}]
&=& [r/\Delta^{1/2}, \Delta^{1/2} , r^2]\ .
\end{eqnarray}

The orthonormalization of the coordinate frame in two steps (diagonalization and normalization) 
$\omega^{{\hat\alpha}}= \omega^{{\hat\alpha}}{}_\beta \rmd x^{\beta}$, 
$e_{{\hat\alpha}}= e^\beta{}_{{\hat\alpha}}\partial_\beta$
defines two matrices
\beq
(\omega^{{\hat\alpha}}{}_\beta)
= \left(\begin{array}{cccc} 1&0&0&0\\ 0&1&0&0\\ 0&0&1&0\\ -\nu_{(n)}&0&0&1\end{array}\right)
  \left(\begin{array}{cccc} N&0&0&0\\ 0&g _{rr}^{1/2}&0&0\\ 0&0&g _{\theta\theta}^{1/2}&0\\                         
                    0&0&0&g _{\phi\phi}^{1/2}\end{array}\right)
\ ,
\eeq
where now 
\beq
\nu_{(n)} = \nu(n,m) = -\nu(m,n) = -g _{\phi\phi}^{1/2} N^{-1}N^\phi
  = 2aM/(r\Delta^{1/2})
\eeq
is the sign-reversed relative velocity with respect to the ZAMOs of the $t$ coordinate lines (static observer world lines).
Its derivative can be expressed in terms of the single nonvanishing component of the ZAMO shear
\begin{eqnarray}
2\theta(n)_{\hat r\hat\phi}
&=& -g_{\hat\phi\hat\phi}^{-1/2} N^{-1} N^\phi{}_{,{\hat r}}
= \nu_{(n)}{}_{,\hat r} + \nu_{(n)} a(n)^{\hat r} + \nu_{(n)} \kappa(\phi,n)^{\hat r}
\end{eqnarray}
whose sign-reversed value plays the role of the vorticity component $\omega(m)^{\hat\theta}$ in the previous discussion.

Similarly one can introduce the slicing orthonormal frame $ W^{{\hat\alpha}} = W^{{\hat\alpha}}{}_B \rmd X^B$ associated with the Fermi coordinates, i.e., with the timelike axis orthogonal to the $T$ coordinate lines,  
with (to first order)
\beq\fl\qquad
(W^{{\hat\alpha}}{}_B) 
= \left(\begin{array}{cccc} 1&0&0&0\\ \Omega X^3&1&0&0\\ 0&0&1&0\\ -\Omega X^1 &0&0&1\end{array}\right)
  \left(\begin{array}{cccc} 1+AX^1&0&0&0\\ 0&1&0&0\\ 0&0&1&0\\  0&0&0&1\\ 
   \end{array}\right)\ .
\eeq

Again fixing the lowest order coefficients as before
\beq
 \zero \omega^{{\hat\alpha}}{}_\beta (x^\alpha - x^\alpha_0) 
 = \delta^{{\hat\alpha}}{}_B X^B +  b^{{\hat\alpha}}_{BC} X^B X^C  + O(3)\ ,
\eeq
then
\beq
 \zero \omega^{{\hat\alpha}} 
 =  [\delta^{{\hat\alpha}}{}_B + 2 b^{\hat\alpha}_{BC} X^C  ] \rmd X^B  + O(2)\ .
\eeq

The $\hat r$ derivatives are now
\beq
  (\omega^{\hat\alpha}{}_{\gamma,{\hat r}}  e^\gamma{}_{\hat\beta})
 = {\rm diag}(a(n)_{\hat r}, -\kappa(x^i,n)_{\hat r})
    - (2\theta(n)_{\hat r\hat\phi} \delta^{\hat\alpha}{}_0  \delta^{\hat\phi}{}_{\hat\beta}) 
      \ ,
\eeq
where $a(n)^{\hat r}=(\ln N)_{\hat r}$ is the radial acceleration, 
$\theta(n)_{\hat r\hat\phi}$ is the shear (since the expansion scalar is zero), 
and $\kappa(x^i,n)^{\hat r}=-(\ln  g_{ii}^{1/2})_{,\hat r}$ are the radial components of the curvature vectors associated with the diagonal metric coefficients
\beq
 a(n)^{\hat r} = \frac{\mathcal{M}\Delta^{1/2}}{r^2(r-2\mathcal{M})}\ ,\quad
 \theta(n)_{\hat r\hat\phi} = -\frac{aM(3r^2+a^2)}{r^2(r^3+a^2r+2a^2M)}\ ,\
\eeq
\beq\fl
 \kappa(r,n)^{\hat r} = \frac{Mr-a^2}{r^2\sqrt{\Delta}}\ ,\
 \kappa(\theta,n)^{\hat r} = -\frac{\sqrt{\Delta}}{r^2}\ ,\
 \kappa(\phi,n)^{\hat r} =-\frac{\sqrt{\Delta}(r^3-a^2M)}{r^2(r^3+a^2r+2a^2M)}\ .
\eeq
However, although it has zero vorticity,  the ZAMO observer congruence is not a Killing congruence and the Fermi-Walker angular velocity of its frame is instead given by the shear component $\Omega=\theta_{\hat r\hat\phi}$. This is a consequence of the fact that in the equatorial plane, the ZAMO acceleration is radial so its 
Frenet-Serret frame is locked to the spherical coordinate frame and hence the Frenet-Serret angular velocity coincides with the Fermi-Walker angular velocity which in this case reduces to the first torsion $\tau_1(n)$ alone (up to sign). This may be calculated using Eqs.~(3.1), (3.11), and (6.1) of reference \cite{circfs}, the latter of which in the equatorial plane has only one component, so when the acceleration is radially outward one has $0\leq\kappa = a(n)^{\hat r}$ (see the comment at the end of the paragraph containing this equation).
For that case the third Frenet-Serret frame vector is aligned with $-e_{\hat\theta}$, so $\tau_1(n) = -\Omega$.

Thus with $A= a(n)^{\hat r}$ and $\Omega = \theta(n)_{\hat r\hat\phi}$ either assumed (based on the previous reasoning) or determined from the first order terms by solving the linear system of equations, one finds
\begin{eqnarray}
\label{trasf}
\fl \qquad
\phantom{---\zero M_\phi (\phi-\phi_0)}\zero N(t-t_0)&=& T 
    -\zero \theta(n)_{\hat r\hat\phi} X^1 X^3
     + O(3)
\ , \nonumber\\\fl \qquad
\phantom{I-\zero M_\phi (\phi-\phi_0)}\zero g_{rr}^{1/2}(r-r_0)
&=& X^1(1+ {\textstyle \frac12}\zero \kappa(r,n)^{\hat r}X^1)
          -{\textstyle \frac12} \zero \kappa(\theta,n)^{\hat r} (X^2)^2 
\ ,\nonumber \\\fl \qquad &&
          -{\textstyle \frac12} \zero \kappa(\phi,n)^{\hat r} (X^3)^2
           + O(3)
\ ,\nonumber \\\fl \qquad
\phantom{I-\zero M_\phi (\phi-\phi_0)}\zero g_{\theta\theta}^{1/2}(\theta-\theta_0) 
&=& X^2(1+ \zero \kappa(\theta,n)^{\hat r}X^1)
 + O(3)
\ ,\nonumber \\\fl \qquad
\zero g_{\phi\phi}^{1/2}[(\phi-\phi_0) +\, \zero N^\phi (t-t_0)]
&=& X^3(1+ \zero \kappa(\phi,n)^{\hat r}X^1)
 + O(3)
\ ,
\end{eqnarray} 
where again all the Boyer-Lindquist metric quantities are evaluated at $X^i=0$ (namely $r=r_0$, $\theta=\theta_0=\pi/2$). Here $X^3=0$ requires that $\phi=\phi_0 -\zero N^\phi (t-t_0)$, which describes how the azimuthal coordinate changes along a ZAMO world line.
The final coordinate transformation is then time-dependent as expected
\begin{eqnarray}
\fl \quad
t-t_0&=& \zero N^{-1}[T 
    -\zero \theta(n)_{\hat r\hat\phi} X^1 X^3]
     + O(3)
\ , \nonumber\\\fl \quad
r-r_0
&=& \zero g_{rr}^{-1/2}[X^1(1+ {\textstyle \frac12}\zero \kappa(r,n)^{\hat r}X^1)
\ ,\nonumber \\\fl \quad &&
          -{\textstyle \frac12} \zero \kappa(\theta,n)^{\hat r} (X^2)^2 
          -{\textstyle \frac12} \zero \kappa(\phi,n)^{\hat r} (X^3)^2 ]
           + O(3)
\ ,\nonumber \\\fl \quad
\theta-\theta_0 
&=& \zero g_{\theta\theta}^{-1/2} [ X^2(1+ \zero \kappa(\theta,n)^{\hat r}X^1) ]
 + O(3)
\ , \\\fl \quad
\phi-\phi_0
&=& \zero g_{\phi\phi}^{-1/2} [X^3(1+ \zero \kappa(\phi,n)^{\hat r}X^1) 
    -\zero \nu(m,n) (T -\zero \theta(n)_{\hat r\hat\phi} X^1 X^3)  ]
 + O(3)\nonumber
\ .
\end{eqnarray}
in order for $\phi$ to increase along the ZAMO world lines, which are moving with respect to the Boyer-Lindquist coordinates.
The first equation shows the second order desynchronization of the local Fermi time coordinate relative to the $t$ coordinate hypersurface due to the quadratic deviation of the time hypersurface from the locally geodesic $T$ coordinate hypersurface due to its extrinsic curvature (in this case $K_{\alpha\beta}=-\theta_{\alpha\beta}$).

\section{Comoving coordinates for any circular orbit}

The calculation of the static observer case is valid for any uniformly rotating Killing observer congruence provided one uses comoving coordinates for the given circular orbit under consideration, an idea elegantly exploited first by Rindler and Perlick \cite{rindlerperlick,rug} to study gyroscopic precession.
These comoving coordinates are related to the usual Boyer-Lindquist coordinates by
\beq
\tilde t = t\ ,\
\tilde r = r\ ,\
\tilde\theta = \theta\ ,\
\tilde\phi = \phi-\zeta t\ ,
\eeq
where $\zeta$ is the constant angular velocity of the new circularly rotating observer parametrizing its 4-velocity
$\tilde m=\Gamma (\partial_t+\zeta \partial_\phi)$, where $\Gamma>0$ is a normalization factor.
All of the results of section 2 (except for the explicit Boyer-Lindquist coordinate expressions for geometric quantities) are valid with tildes referring to such a coordinate system with $m\to \tilde m =U$. One can then relate all the tilde metric quantities back to the original static observer quantities to evaluate the transformation to Fermi coordinates along a general timelike circular orbit in the equatorial plane. The details of these calculations involving the transformation laws with some of the machinery of \cite{idcf2}
would  lead us a bit astray from the main point of the present article. However, one should note
that for the extremely accelerated observers, the Fermi-Walker angular velocity is zero, so the $\tilde\Omega$ term drops out of the transformation, while for the circular geodesics, the acceleration term is zero and drops out of the Fermi metric.

We limit ourselves to sketching the results.  Let $U$ denote the 4-velocity of the circular orbit, in relative motion with constant velocity $\nu$ with respect to $m$
\beq
U=\gamma [\, m +\nu \, \omega^{\hat \phi}\, ]
\ ,\qquad  
\gamma=(1-\nu^2)^{-1/2}\ ,
\eeq
related to the coordinate angular velocity $\zeta$ by the identification $U=\Gamma[\partial_t +\zeta \partial_\phi]$ which leads to 
\beq
  \zeta = M \nu /[\gamma_{\phi\phi}^{1/2}+ M M_\phi \nu]\ .
\eeq
The natural Lie dragged frame naturally defined all along this orbit is obtained by boosting the one adapted to the threading congruence in the azimuthal direction
\begin{eqnarray}
\fl\qquad
&&
W^{\hat t} = \gamma [\,\omega^{\hat t} - \nu\omega^{\hat \phi}\, ] \ , \quad
W^{\hat r}  = \omega^{\hat r} = g_{rr}^{1/2}\, \rmd r \ , \quad
W^{\hat \theta} = \omega^{\hat \theta} = g_{\theta\theta}^{1/2}\, \rmd \theta\ ,
\nonumber\\
\fl\qquad
&&  
W^{\hat \phi} = \gamma [-\nu \omega^{\hat t} +\omega^{\hat \phi}\, ]\ .
\end{eqnarray}
The relationship between the Boyer-Lindquist coordinates ($t, r, \theta, \phi$) and the Fermi coordinates ($T,X^i$) in the neighborhood of the circular orbit world line has the same form as before but with two transformed coefficients for the Fermi-Walker angular velocity and the azimuthal curvature on the right hand side and the change of azimuthal coordinate on the left hand side
\begin{eqnarray}
\fl
&&DT  = 
T - \Omega(\nu) X^1X^3  + O(3)
\ , \nonumber\\ \fl 
&& DR
=X^1(1
 + {\textstyle \frac12} \zero \kappa(r,m)^{\hat r}X^1) 
 - {\textstyle \frac12} \zero \kappa(\theta,m)^{\hat r} (X^2)^2 
 - {\textstyle \frac12} \zero \kappa(\phi,U)^{\hat r} (X^3)^2 + O(3)
\ ,\nonumber \\\fl
&& D\Theta 
=X^2(1+ \zero \kappa(\theta,m)^{\hat r}X^1)
 + O(3)
\ ,\nonumber \\\fl
&& D\Phi
= X^3(1+\zero \kappa(\phi,U)^{\hat r} X^1) 
 + O(3)
\ ,
\end{eqnarray} 
where 
\begin{eqnarray}
\fl\quad
DT &=& \gamma \left[\zero M(t-t_0) 
            -(\zero M \zero M_\phi +\nu \zero \gamma_{\phi\phi}^{1/2} )(\phi-\phi_0)\right]\ , \nonumber \\
\fl\quad
DR &=& \zero \gamma_{rr}^{1/2}(r-r_0)\ , \nonumber \\
\fl\quad
D \Theta &=& \zero \gamma_{\theta\theta}^{1/2}(\theta-\theta_0)\ , \nonumber \\
\fl\quad
D \Phi &=& \gamma \left[-\zero M\nu (t-t_0)
             +(\zero M \zero M_\phi \nu + \zero \gamma_{\phi\phi}^{1/2} )(\phi-\phi_0)\right]\ .
\end{eqnarray}
The condition $X^3=0$ for the origin of the coordinates reproduces the equation for the orbit
$\phi-\phi_0 = \zeta (t -t_0)$.

The new 4-acceleration is now
\beq
\kappa (\nu) = \gamma^2 \zero \kappa(\phi,m)^{\hat r}  (\nu-\nu_+)(\nu-\nu_-)
\equiv A \ ,
\eeq
where the geodesic velocities are
\beq
  \nu_\pm = \Delta^{1/2}/[a\pm (r-2\mathcal{M})(r/\mathcal{M})^{1/2} ]\ ,
\eeq
while
\beq
\Omega(\nu) = {\textstyle \frac12} \gamma^{-2} \rmd \kappa(\nu)/ \rmd \nu
\eeq
is the Fermi-Walker angular velocity of the new frame and 
\beq
 \kappa(\phi,U)^{\hat r} = -\gamma^2  \kappa(\phi,m)^{\hat r}(1-\nu\nu_+)(1-\nu\nu_-)
  = - \kappa(\nu^{-1})
\eeq
is the comoving azimuthal curvature.
The following relations among the geodesic velocities and kinematical fields have been used \cite{idcf2}
\beq
a(m)^{\hat r}= \kappa(\phi,m)^{\hat r} \nu_+ \nu_-\ , \qquad 
-2\omega(m)^{\hat\theta} = \kappa(\phi,m)^{\hat r}( \nu_+ + \nu_-)
\eeq
in obtaining these formulas.

For the most interesting case of circular geodesics (zero acceleration) where $\nu =\nu_\pm$, the azimuthal curvature reduces to
\beq
   \kappa(\phi,U_\pm)^{\hat r} =  \kappa(\phi,m)^{\hat r} - a(m)^{\hat r}
    = \kappa(\phi,m)^{\hat r}(1-\nu_+\nu_-)
\eeq
while the frame angular velocity takes the simple form 
$ \Omega(\nu_\pm)= 
\mp (\mathcal{M}/r^3)^{1/2}$ in contrast with the orbital angular velocity 
$\zeta_\pm = \pm(\mathcal{M}/r^3)^{1/2}/[1\pm a(\mathcal{M}/r^3)^{1/2}] $. In the Schwarzschild case, these are just differing by sign but they are rates taken with respect to the proper time and coordinate time respectively, so they do not cancel each other out with respect to axes whose  directions  are fixed at spatial infinity, leading to the known precession of spin vectors along these orbits.

On the other hand for the extremely accelerated world lines \cite{def,sem,idcf2}, the frame angular velocity vanishes  because the Lie dragged frame coincides with the Fermi-Walker transported frame along the world line so that $DT = T +O(3)$. Finally the azimuthal curvature $ \kappa(\phi,U)^{\hat r}$ is zero for the so called \lq\lq Lie relatively straight" spacelike circular orbits which move in the direction orthogonal to the timelike geodesics with relative velocity $ \nu=\nu_\pm^{-1}$, so that $D\phi = X^3 + O(3)$.

\section{Third order coordinate transformation for static observers}

Given the coordinate transformation from Boyer-Lindquist coordinates to Fermi coordinates up to second order for the static observers, one can go to the next order with the same strategy. The components of the Fermi metric up to second order were given by Ni and Zimmerman, following the same conventions as Misner, Thorne and Wheeler but including the extra acceleration and angular velocity terms. In our case with only the orthonormal components $a^1 =A$ and $\omega^2=\Omega$ nonvanishing we have
\begin{eqnarray}\fl
\rmd s^2
&=&
-\left\{ (1+AX^1)^2-\Omega^2 [(X^1)^2+(X^3)^2]+E_{\hat r \hat r}(X^1)^2+E_{\hat \theta \hat \theta}(X^2)^2+E_{\hat \phi \hat\phi}(X^3)^2
\right\} 
\rmd T^2\nonumber\\
\fl &&
+2\left\{ (\Omega X^3-{\textstyle\frac23} H_{\hat r \hat \theta}X^1 X^3) \rmd X^1+{\textstyle\frac23} H_{\hat r \hat \theta}X^2 X^3\rmd X^2\right.
\nonumber\\\fl &&
\left.+\{- \Omega X^1+{\textstyle\frac23}H_{\hat r \hat \theta}[(X^1)^2-(X^2)^2]\} \rmd X^3
\right\} 
\rmd T
\nonumber\\\fl &&
+\left\{
1+{\textstyle\frac13} [E_{\hat \phi \hat \phi}(X^2)^2+E_{\hat \theta \hat\theta}(X^3)^2]
\right\} (\rmd X^1)^2
-{\textstyle\frac23} E_{\hat r \hat r}X^2 X^3 \rmd X^2 \rmd X^3
\nonumber\\\fl &&
+ \left\{
1+{\textstyle\frac13} [E_{\hat \phi \hat \phi}(X^1)^2+E_{\hat r \hat r}(X^3)^2]
\right\} (\rmd X^2)^2
-{\textstyle\frac23} E_{\hat \theta \hat \theta}X^1 X^3  \rmd X^1 \rmd X^3
\nonumber\\\fl &&
+\left\{
1+{\textstyle\frac13} [E_{\hat \theta \hat\theta}(X^1)^2+E_{\hat r \hat r}(X^2)^2]
\right\} (\rmd X^3)^2
-{\textstyle\frac23} E_{\hat \phi \hat \phi}X^1 X^2  \rmd X^1 \rmd X^2\ .
\end{eqnarray}

Here the only nonvanishing independent components of the Riemann tensor are the following electric and magnetic parts of the Weyl tensor
\begin{eqnarray}\fl\quad
 &&
 E_{\hat r\hat r} = -\frac{\mathcal{M}}{r^3}\frac{2\Delta+a^2}{\Delta-a^2}
 = R_{\hat0\hat r\hat0\hat r} = -R_{\hat0\hat r\hat0\hat r}\ ,\
 E_{\hat\theta\hat\theta} = \frac{\mathcal{M}}{r^3}\frac{\Delta+2a^2}{\Delta-a^2}
 = R_{\hat0\hat\theta\hat0\hat\theta} = -R_{\hat r\hat\phi\hat r\hat\phi}\ ,
 \nonumber\\ \fl\quad
 &&
  E_{\hat\phi\hat\phi} = \frac{\mathcal{M}}{r^3}
 = R_{\hat0\hat\phi\hat0\hat\phi} = -R_{\hat r\hat\theta\hat r\hat\theta}\ ,\
 H_{\hat r\hat\theta} = -\frac{3\mathcal{M}}{r^3}\frac{\Delta^{1/2} a}{\Delta-a^2}
 = R_{\hat0\hat r\hat r\hat\phi}
 \ .
\end{eqnarray}

One then finds the additional terms in the corresponding coordinate transformation in an abbreviated notation
\begin{eqnarray}
\label{trasfM2}
\fl&& 
M[(t-t_0)- M_\phi (\phi-\phi_0)] = \ldots\nonumber\\
\fl&& 
+ \frac{X^3}{3}\bigg[
(X^1)^2\left(H_{\hat r \hat \theta}-3\omega^{\hat \theta}(\kappa_{\phi}{}^{\hat r} -a^{\hat r})\right)
-(X^2)^2H_{\hat r \hat \theta}
+(X^3)^2\kappa_{\phi}{}^{\hat r}\omega^{\hat \theta}
\bigg]+O(4)\ ,
\nonumber\\
\fl&&
\gamma_{rr}^{1/2}(r-r_0)= \ldots\nonumber\\
\fl&& 
+ \frac{X^1}{6}\bigg[
(X^1)^2(\partial_{\hat r}\kappa_{r}{}^{\hat r}+(\kappa_{r}{}^{\hat r})^2)
-(X^2)^2(3\partial_{\hat r}\kappa_{\theta}{}^{\hat r}+3\kappa_{r}{}^{\hat r}\kappa_{\theta}{}^{\hat r}+2(\omega^{\hat \theta})^2-2a^{\hat r}\kappa_{\phi}{}^{\hat r})
\nonumber\\
\fl&& 
-(X^3)^2(\partial_{\hat r}\kappa_{\phi}{}^{\hat r}+2(\kappa_{\phi}{}^{\hat r})^2+3\kappa_{\phi}{}^{\hat r}\kappa_{r}{}^{\hat r})
\bigg]+O(4)\ ,
\nonumber \\
\fl&&
\gamma_{\theta\theta}^{1/2}(\theta-\theta_0)= \ldots\nonumber\\
\fl&& 
+ \frac{X^2}{6}\bigg[
(X^1)^2((\omega^{\hat \theta})^2-a^{\hat r}\kappa_{\phi}{}^{\hat r}+3\partial_{\hat r}\kappa_{\theta}{}^{\hat r}+3(\kappa_{\theta}{}^{\hat r})^2) 
\nonumber\\
\fl&& 
-(X^2)^2((\omega^{\hat \theta})^2-a^{\hat r}\kappa_{\phi}{}^{\hat r}+\partial_{\hat r}\kappa_{\theta}{}^{\hat r}+(\kappa_{\theta}{}^{\hat r})^2)
-(X^3)^2(\partial_{\hat \theta}\kappa_{\phi}{}^{\hat \theta}+2\kappa_{\phi}{}^{\hat r}\kappa_{\theta}{}^{\hat r})
\bigg]+O(4)\ ,
\nonumber \\
\fl&&
\gamma_{\phi\phi}^{1/2}(\phi-\phi_0)= \ldots\nonumber\\
\fl&& 
+ \frac{X^3}{3}\bigg[
(X^1)^2(\partial_{\hat r}\kappa_{\phi}{}^{\hat r}+2(\kappa_{\phi}{}^{\hat r})^2)
+(X^2)^2(\partial_{\hat \theta}\kappa_{\phi}{}^{\hat \theta}-\kappa_{\phi}{}^{\hat r}\kappa_{\theta}{}^{\hat r})
\nonumber \\
\fl&&
-(X^3)^2(\kappa_{\phi}{}^{\hat r})^2
\bigg]+O(4)\ ,
\end{eqnarray}
where it is understood that all the Boyer-Lindquist metric quantities are evaluated at $r=r_0$, $\theta=\theta_0=\pi/2 $ and one has the explicit formulas
\begin{eqnarray}
\partial_{\hat r}\kappa_{r}{}^{\hat r}&=&\frac{1}{r ^2}\left[\frac{2a^2}{r ^2}+\left(1-\frac{2{\mathcal M}}{r }\right)-\frac{(r -{\mathcal M})^2}{\Delta }\right]\ , \nonumber\\
\partial_{\hat r}\kappa_{\theta}{}^{\hat r}&=&\frac{\Delta +(a^2-{\mathcal M}r )}{r ^4}\ , \nonumber\\
\partial_{\hat r}\kappa_{\phi}{}^{\hat r}&=&\frac{1}{r ^2}\left[\frac{{\mathcal M}a^2(4{\mathcal M}-3r )}{r ^4(1-2{\mathcal M}/r )^2}+\frac{(5{\mathcal M}-3r ){\mathcal M}}{r ^2(1-2{\mathcal M}/r )}+\frac{(r -{\mathcal M})^2}{\Delta}\right]\ , \nonumber\\
\partial_{\hat \theta}\kappa_{r}{}^{\hat \theta}&=&-\frac{a^2}{r ^4}=\partial_{\hat \theta}\kappa_{\theta}{}^{\hat \theta}\ , \nonumber\\
\partial_{\hat \theta}\kappa_{\phi}{}^{\hat \theta}&=&\frac{1}{r ^4}\left[\frac{a^2}{1-2{\mathcal M}/r }+r ^2-a^2\right]\ ,
\end{eqnarray}
in terms of which the electric part of the Riemann tensor consists of the following combinations
\begin{eqnarray}	
E_{\hat r \hat r}&=&-\partial_{\hat \theta}\kappa_{\phi}{}^{\hat \theta}+\kappa_{\phi}{}^{\hat r}\kappa_{\theta}{}^{\hat r}\ , \nonumber\\
E_{\hat \theta \hat \theta}&=&-\partial_{\hat r}\kappa_{\phi}{}^{\hat r}-3(\omega^{\hat \theta})^2+(\kappa_{\phi}{}^{\hat r})^2\ , \nonumber\\
E_{\hat \phi \hat \phi}&=&-\partial_{\hat r}\kappa_{\theta}{}^{\hat r}-\partial_{\hat \theta}\kappa_{\theta}{}^{\hat \theta}+(\kappa_{\theta}{}^{\hat r})^2=(\omega^{\hat \theta})^2-a^{\hat r}\kappa_{\phi}{}^{\hat r}\ ,
\end{eqnarray}
while another relation between these quantities follows from $ E^{\hat i}{}_{\hat i}=0$.

For the Schwarzschild case, this simplifies considerably.
One has 
\begin{eqnarray}\fl
&&
\partial_{\hat r}\kappa_{r}{}^{\hat r}+(\kappa_{r}{}^{\hat r})^2+2\frac{\mathcal M}{r ^3}=0\ ,\ 
\partial_{\hat r}\kappa_{\theta}{}^{\hat r}-(\kappa_{\theta}{}^{\hat r})^2+\frac{\mathcal M}{r ^3}=0\ ,\
\partial_{\hat r}\kappa_{\phi}{}^{\hat r}-(\kappa_{\phi}{}^{\hat r})^2+\frac{\mathcal M}{r ^3}=0\ ,\
 \nonumber\\
\fl&&
\partial_{\hat \theta}\kappa_{r}{}^{\hat \theta}=0\ ,\  
\partial_{\hat \theta}\kappa_{\phi}{}^{\hat \theta}-(\kappa_{\phi}{}^{\hat r})^2-2\frac{\mathcal M}{r ^3}=0\ ,
\end{eqnarray}
where we have used the relations
\begin{eqnarray}\fl&&
\kappa_{\phi}{}^{\hat r} = \kappa_{\theta}{}^{\hat r}
=-\frac{1}{r }\left(1-2\mathcal M/r \right)^{1/2}\ , \quad 
\kappa_{\phi}{}^{\hat r}\kappa_{r}{}^{\hat r}
=-\frac{\mathcal M}{r ^3}\ ,
\nonumber\\&&
a^{\hat r}=\kappa_{r}{}^{\hat r}
=\frac{\mathcal M}{r ^2}\left(1-2\mathcal M/r \right)^{-1/2}\ .
\end{eqnarray}
The components of the electric and magnetic parts of the Riemann tensor reduce to 
\beq
[E_{\hat r \hat r}, E_{\hat \theta \hat \theta}, E_{\hat \phi \hat \phi}, H_{\hat r \hat \theta}]=\frac{\mathcal M}{r ^3}[-2, 1, 1, 0]\ .
\eeq
The corresponding coordinate transformation is then given in terms of only two independent coefficients (the radial and tangential intrinsic curvatures) whose product is the single spacetime curvature invariant
\begin{eqnarray}
\fl 
M(t-t_0) &=& T +O(4)\ ,
\nonumber\\
\fl
g_{rr}^{1/2}(r-r_0)&=& X^1
+ {\textstyle\frac12} \left[\kappa_{r}{}^{\hat r}(X^1)^2-\kappa_{\phi}{}^{\hat r}[(X^2)^2+(X^3)^2]\right] 
\nonumber\\
\fl&&
+ {\textstyle\frac13} X^1 \left[\kappa_{r}{}^{\hat r}\kappa_{\phi}{}^{\hat r}(X^1)^2 
    -{\textstyle\frac12}\kappa_{\phi}{}^{\hat r}[(X^2)^2+(X^3)^2]
            (3\kappa_{\phi}{}^{\hat r} + 4\kappa_{r}{}^{\hat r}) \right]
    +O(4)\ ,
\nonumber \\
\fl
g_{\theta\theta}^{1/2}(\theta-\theta_0)&=& X^2
+ \kappa_{\phi}{}^{\hat r}X^1X^2
+ {\textstyle\frac13}X^2 \bigg[\kappa_{\phi}{}^{\hat r}(3\kappa_{\phi}{}^{\hat r}+\kappa_{r}{}^{\hat r})(X^1)^2
\nonumber\\
\fl&&
-(\kappa_{\phi}{}^{\hat r})^2(X^2)^2
+{\textstyle\frac12} \kappa_{\phi}{}^{\hat r}(-3\kappa_{\phi}{}^{\hat r}+2\kappa_{r}{}^{\hat r})(X^3)^2\bigg] 
+O(4)\ ,
\nonumber \\
\fl
g_{\phi\phi}^{1/2}(\phi-\phi_0)&=& X^3+\kappa_{\phi}{}^{\hat r}X^1X^3
+{\textstyle\frac13} X^3 \bigg[\kappa_{\phi}{}^{\hat r}(3\kappa_{\phi}{}^{\hat r}+\kappa_{r}{}^{\hat r})(X^1)^2
\nonumber\\
\fl&&
-2\kappa_{\phi}{}^{\hat r}\kappa_{r}{}^{\hat r}(X^2)^2-(\kappa_{\phi}{}^{\hat r})^2(X^3)^2\bigg]
+O(4)\ ,
\end{eqnarray}
where again all the coefficients are understood to be evaluated at $r=r_0,\theta=\theta_0=\pi/2$.

\section{Concluding remarks}

We have derived the local coordinate transformation between Boyer-Lindquist coordinates and Fermi coordinates (up to second order in the Fermi coordinates) adapted to an accelerated observer located at a fixed point of the equatorial plane of the Kerr solution, and extended this result to any uniformly circularly rotating observer in that plane, including the ZAMOs, where the transformation may be re-expressed more appropriately in terms of the ZAMO geometrical quantities.  For the static observers the result extends to the third order in the Fermi coordinates but involves many coefficients from which the Weyl curvature tensor components can be formed. In the Schwarzschild case the situation simplifies considerably, 
generalizing the explicit first order expressions obtained by Linet and giving a geometric interpretation to the contributing terms. Since Fermi coordinates describe the most suitable apparatus of clocks and rods for test observers, this coordinate transformation may be useful in anchoring local Fermi coordinate calculations to the Kerr spacetime manifold.

\end{document}